# Accuracy of Discrete Markov Approximation in the Problems of Estimation of Random Field Characteristics

U.N. Brimkulov, Ch.N. Jumabaeva, K. Baryktabasov

Kyrgyz Turkish Manas University, Bishkek, Kyrgyzstan

*Abstract.* **The covariance matrix of measurements of Markov random fields (processes) has useful properties that allow to develop effective computational algorithms for many problems in the study of Markov fields on the basis of field observations (parametric identification problems, filtering problems, interpolation problems and others). Therefore, approximation of arbitrary random fields by Markov fields is of great interest, as it gives an opportunity to use computationally efficient algorithms of Markov fields analysis to study them.**

**The paper deals with approximation of the covariance matrix of the field being observed with the help of covariance matrix of a multiply connected (*m*-connected) Markov field. Using computer simulation, the accuracy of such replacements at different values of the connectivity coefficient *m* for the problem of parametric identification of deterministic component of the field has been studied. Various models of deterministic polynomial component of the field and a number of covariance functions that are often used as a mathematical model of real random noise and measurements noise have been reviewed.**

**It has been shown that for many problems such approximation, even at small connectivity *m* values of approximating Markov field, provides necessary accuracy. This allows to achieve good compromise between accuracy of the estimates and complexity of calculating them.**

*Index Terms.* **Multiply connected (*m*-connected) Markov field (Markov field of *m*-th order), covariance matrix of the measurement, parametric identification of the random field, Generalized Least Squares Estimates (GLSE) method, Best Linear Unbiased Estimates (BLUE), Discrete Markov Approximation (DMA), approximation accuracy, computer simulation.**

## I. INTRODUCTION

PROBLEMS of study of random fields and processes could be found in many applications, for example, in the study of spatial and temporal variability of oceanographic fields (fields of flow velocities, fields of temperature and sea surface level, sea water density and salinity fields , etc.), in problems of statistical radio engineering and image modeling, as well as in many other engineering tasks.

Method of least squares (LSE), weighted or generalized LSE (WLSE, GLSE) methods, Kalman-Bucy filter and other techniques are commonly used in estimation, filtering and interpolation problems of random fields based on field measurements. When physical random fields being studied are Markov random fields in a wide sense (in other words, Gauss-Markov random fields), or they can be approximated by Markov fields, we come to the computational schemes where we can meet the covariance matrices of measurements, inversions of which have diagonal structure:

a) tridiagonal structure, in case of a simple (one-dimensional, ordinary connected) Markov field;

b) band-diagonal structure for m-connected Markov field (in literature they are also called the Gauss-Markov fields of m-th order (M-th order Gauss-Markov field);

c) block-tridiagonal structure, in the case of vector ordinary connected Markov field;

d) block-band-diagonal structure, in the case of m-connected vector Markov field.

A lot of papers are dedicated to the studies of matrices with diagonal structure, for example [1-5, 13, 19-21]. We should especially single out the most general article by G. Meurant, 1992 [19], which represents in-depth review and analysis of the results of the study of the properties of inversions in respect of symmetric tridiagonal, block-tridiagonal and band matrices. The article summarizes a lot of results and gives full enough (34 titles) bibliography of publications in this field, in the period of 1944 - 1992. The review starts with the first publication by D. Moskovitz, 1944 [20], which gives analytical expressions for the problems, considering 1D and 2D Poisson models.

Matrices connectivity, inversion of which have diagonal structure with Gauss-Markov Markov processes, were studied in works [1-5, 15-17]. In work [1], devoted to the assimilation of data during study of large, multi-dimensional, time-dependent fields by taking into account the structure of the matrix of measurements, it was managed to develop 4 efficient algorithms of Kalman-Bucy filter, reducing computing costs

U.N.Brimkulov, is with the Computer Engineering Department, Engineering Faculty, Kyrgyz-Turkish Manas University, 56 Mira Avenue, Bishkek 720042, Kyrgyzstan (e-mail: ulan.brimkulov@manas.edu.kg; unbrim@gmail.com)

Ch.N.Jumabaeva, is with the Computer Engineering Department, Engineering Faculty, Kyrgyz-Turkish Manas University, 56 Mira Avenue, Bishkek 720042, Kyrgyzstan (e-mail: chinara.jumabaeva@manas.edu.kg)

K.K.Baryktabasov, is with the Computer Engineering Department, Engineering Faculty, Kyrgyz-Turkish Manas University, 56 Mira Avenue, Bishkek 720042, Kyrgyzstan (kasym.baryktabasov@manas.edu.kg)

by 2 orders in comparison with known algorithms. This possibility appears in the case where the measurement errors are approximated by Markov random field. In this case, the inverse of covariance matrix of field measurement errors has a band structure that allows to develop efficient algorithms in the sense of computing. Taking into account the sparseness of measurements, typical for the problems, reviewed in this article (e.g., results of satellite scanning), makes it possible to obtain algorithms that are even more effective.

Matrices, inversion of which are band like, are considered in work [2]. In the paper, tridiagonal matrix is represented as the Hadamard product of three matrices. This leads to a very interesting result, when a random Gauss-Markov process is represented as the product of three independent processes: direct and inverse processes with independent increments and a process with stationary dispersion. Here we can see the connection between the matrices appearing in the expansion of tridiagonal matrix and the processes occurring in the decomposition of a random Gauss-Markov process. In this sense, the positively defined symmetric matrices with band inversions could be considered as a form of representation of Gauss-Markov random processes.

In paper [3] there were obtained algorithms of inversion of L - block band matrices, inversions of which also are L - block band ones. These inversion algorithms are applicable to problems of signal processing when Kalman-Bucy filter (KBF) is used. At the same time, these covariance matrices are approximated by block-band matrix. This gives an opportunity to reduce the computational complexity of the algorithms by 2 order and make KBF algorithms feasible for solving large-scale problems.

In paper [19] there is a reference to the work of Barret, 1979 [21], which introduced the concept of "triangle property" (matrix $R$ has a "triangle property" if $R_{ij} = (R_{ik}R_{kj})/R_{kk}$); matrix, having "triangle property" and non-zero diagonal elements, has tridiagonal inversion and vice versa). It should be noted that "triangle property", introduced in [21], coincides with a discrete form of the condition for the form of the covariance function of Markov process in a wide sense, given in the book by Dub, 1953, Theorem 8.1 [9]. We underline this, as results of Theorem 8.1 serve as a basis to many of the results given in Appendix 1, which are related to the study of the connectivity of matrices, inversions of which are tridiagonal, block, or block-tridiagonal matrices with the covariance matrix of the ordinary (simple) Markov processes, multiply connected Markov processes and vector Markov random processes.

Summarizing content of the above articles it may be noted that all these works study signal processing algorithms, having Markov property. At the same time, the structure of the matrix included in algorithms for processing such signals allows to obtain efficient computational algorithms to solve large-scale problems.

Unfortunately, when dealing with many real problems, we are dealing with the analysis of random processes, not having Markov property. Inversions of covariance matrices of measurements of such processes do not have a diagonal structure and are general type matrices. While processing observations of such processes, it does not give a possibility to use efficient computational schemes, obtained with the structure of diagonal covariance matrices of Markov processes taken into account.

Therefore, there exists an urgent problem of approximating arbitrary random processes by Markov processes. In this regard, works [2,3,5] are of interest, which address the problem of approximating the covariance matrix of a random Gaussian process by covariance matrix, inversion of which is a band matrix. Information loss of such approximation were estimated in [6]. These studies also showed that inversion of these matrices requires knowledge of direct matrix elements only lying inside the band of 2m +1 width.

Questions of non-Markov processes approximation by Markov processes were considered in [7,8,15,16,22-24], as well.

Work [7], dedicated to development of optimal detectors of sequences, presented several new classes of new suboptimal detectors of sequences obtained by replacing the covariance matrix of the process being studied by the covariance matrix of Markov process. Article [8] also considered Markov approximation of non-Markov processes.

Regardless of the results obtained in the works above, we in 1988-1992 obtained the results that overlap with the results of the reviewed papers. Unfortunately, our results are presented only in the form of manuscripts published in Russian (see, e.g. [15]). There is only one article that was translated into English [16] and article[17], which was published recently.

## II. THE PROBLEM OF ESTIMATING CHARACTERISTICS OF A RANDOM FIELD ON THE BASIS OF MEASUREMENT OF ITS REALIZATION

Let us consider one of the well-known problems of the study of random fields - task of finding Linear Unbiased Estimates (LUE) of unknown parameters of the mathematical expectation model (deterministic component) of a random field.

Let the random field being observed is described by the model:

$$Z(t) = \eta(t) + \xi(t) , \qquad (1)$$

where $E\xi(t) = 0, E\xi(s)\xi(t) = k(s,t), \quad t \in T$;

$\eta(t) = \eta(t, \mathbf{B}) = \mathbf{f}^T(t)\mathbf{B}$ - mathematical expectation (deterministic component) of the field described by the linear parameterized model with a vector of known linear-independent functions $\mathbf{f}(t) = \left(f_1(t),\ldots,f_p(t)\right)^T$ and a vector of unknown parameters $\mathbf{B} = \left(B_1,\ldots,B_p\right)^T$; $\xi(t)$ - noise field (interference, measurement noise) with zero mean and known covariance function $k(s,t)$; E - mathematical expectation operator; $T$ - action scope (1).

Let us suppose that the task is set on the basis of discrete measurements $Z(t)$ at the points $T_n = \left\{t_1 < t_n < \ldots < t_n \mid t_i \in T\right\}$ to find the LUE of unknown parameters $\mathbf{B}$ . This problem is solved with the help of GLSE, which are defined by the formula (see, for example, [11,12]):

$$\mathbf{B}_n^{GLSE} = \left[ F_n W_n F_n^T \right]^{-1} F_n W_n \mathbf{Z}_n \qquad (2)$$

in $F_n = [\mathbf{f}(t_1), \dots , \mathbf{f}(t_n)]$ - matrix of $\mathbf{f}(t)$ vector values at the measuring points $T_n$, $W_n = \left[ w_{ij} \right]_{i,j=1}^n$ - the weight positive definite matrix of the size $n \times n$; $\mathbf{Z}_n = \{ Z(t_i) \mid t_i \in T_n \}$ - vector of measurement (observations) of the field at points $T_n$.

The accuracy of estimates (2) is characterized by the covariance (dispersion) matrix defined by formula

$$D_n^{GLSE} = \left[ F_n W_n F_n^T \right]^{-1} F_n W_n K_n W_n^T F_n \left[ F_n W_n F_n^T \right]^{-1}. \qquad (3)$$

where $K_n^{-1}$ - the inversion of the covariance matrix of the vector of measurements $\mathbf{Z}_n$: $K_n = \left\{ k(t_i, t_j) \right\} = \left\{ k_{ij} \right\}$ $(i, j = \overline{1, n})$ $(t_i \in T_n)$.

Highest accuracy of linear GLSE estimates is obtained at [11, 12]

$$W_n = K_n^{-1} \qquad (4)$$

At the same time, GLSE estimates coincide with the BLUE.

Thus, if the covariance matrix $K_n$ is known or can be calculated using functions $k(s,t)$, we can develop the optimal GLSE estimates coinciding with BLUE:

$$\mathbf{B}_n^{BLUE} = \hat{\mathbf{B}}_n = \left[ F_n K_n^{-1} F_n^T \right]^{-1} F_n K_n^{-1} \mathbf{Z}_n, \qquad (5)$$

for the dispersion matrix of which from (3), substituting (4) we obtain

$$D_n^{BLUE} = \left[ F_n W_n F_n^T \right]^{-1}. \qquad (6)$$

Estimates (5) are also called Markov estimates.

If the process $Z$ $(t)$ is Gaussian one, then regardless of the model $\eta(t,\mathbf{B})$ type, the optimal discrete GLSE estimates with a weight matrix $W_n = K_n^{-1}$, are also the maximum likelihood estimates), that is, they maximize the $n$-dimensional likelihood function of measurements $Z_i$ ($i = 1, ..., n$) [18].

If $W_n = I_n$, where $I_n$ - unity matrix, we arrive at the estimates of ordinary LSE.

If $W_n = \Sigma_n$, where $\Sigma_n$ - diagonal matrix, we arrive at WLSE estimates.

Although estimates are optimal (5), usage of them at a large number of measurements becomes difficult or impossible. This is due to the fact that the expressions (5) and (6) include the matrix $K_n^{-1}$, the number of whose elements increases in proportion to the square of the number of dimensions. Computing and storage of matrix $K_n^{-1}$ requires large computational costs (memory required is proportional $n^2$, and the number of operations for matrix inversion is proportional to $n^3$ [14]).

The same matrix $K_n^{-1}$ is part of the expressions to calculate the optimal estimates for more general tasks of processing of random processes on the basis of measurements. For example, the formulas of filtering, interpolation and extrapolation of a random field on the basis of measurements also include matrix $K_n^{-1}$. We do not present here the well-known results, which could be found in numerous papers.

## III. DISCRETE APPROXIMATION OF RANDOM PROCESSES BY $M$-CONNECTED MARKOV PROCESSES

In many cases, we are dealing with observations of random processes, which do not possess the Markov property. Covariance matrix of the measurement of these fields do not have diagonal structure, i.e. they are matrices of general type. It does not allow, while processing observations of such processes, to use efficient computational schemes, typical for Markov fields.

Let us introduce the following notations for covariance matrices of Markov fields:

$K_n^1$ - covariance matrix of simple (singly connected, of the first order) Markov field in a wide sense ($n$ - order of the matrix is equal to the number of field measurements);

$K_n^m$ - covariance matrix of a multiply connected ($m$-connected, $m$-th-order) Markov field ($m$ – half band width is equal to the field connectivity coefficient);

$K_N^m$ - covariance block matrix of a vectorial ($m$-dimensional) Markov field ($N = n \times m$, where $N$ - total dimension of the block matrix, $n$ - number of blocks, $m$ - dimension of the individual blocks).

It would be noted that matrix $K_n^1$ is a special case of matrix $K_n^m$ when $m = 1$.

Here two approaches could be offered, allowing to use results, obtained for Gauss-Markov Markov fields, for processing of observations of non-Markov fields.

1) Approximation of the covariance function of the test field by the covariance function of the Markov field, on the basis of any criterion of proximity of these functions. The problems of approximating other characteristics of a random field could be also considered: distribution law, other momental functions, spectral density, etc. This task for the Kullback-Leibler criterion or criteria of entropy maximum was solved in [2, 23]. Here the best criterion would be one that takes into account the ultimate objective of approximation, that is, for example, in the problems of estimation - the accuracy of the estimates.

2) Approximation of the covariance matrix of measurements of the observed field by covariance matrix of $m$-connected Markov field.

In solving problems of covariant statistics fields on the basis of a finite number of discrete measurements, probabilistic properties of the field are completely determined by dispersion setting at the measurement points and coefficients of the covariance predetrmination between these points. In other words, the covariance matrix of the measurements completely characterizes the probabilistic properties of the covariance field for discrete tasks (see, also similar conclusion in [2]). (Here and below, covariance is used to name random fields, whose probabilistic properties are completely determined by the covariance function of the field.)

*Definition 1. Replacing the covariant matrix of measurements of the first (approximated) field by covariance matrix of the measurements of the second (approximating) field on the basis of predetermined criteria of such replacement is called discrete approximation of one covariant field by another one.*

Let us consider the problem of estimating a random field by means of discrete GLSE (see. Section 2). As already noted, the properties of GLSE estimates strongly depend on the type of weight matrix selected for development of estimates. If evaluated model is linear with respect to the estimated parameters, and matrix $K_n^{-1}$ is used as the weight matrix GLSE, the GLSE estimates coincide with the BLUE (see (4)). The difficulties of calculating GLSE estimates with weight matrix $W_n = K_n^{-1}$ were analyzed above.

Easiness to calculate the GLSE estimates may be taken as one of the criteria for selecting the weighting matrix $W_n$. From this point of view, it is very convenient to use band weight matrices as weight ones. Previously it has been shown that those matrices have the band type with a width of the band $2m+1$, are inverse to covariance measurements matrices of m-connected Markov fields. Thus, if the covariance matrix of measurements of the observed field $K_n$ is replaced by the covariance matrix of measurements of m-connected Markov field $K_n^m$, the weight matrix of GLSE estimates will be band type matrix.

This makes it possible to significantly simplify calculation of GLSE estimates.

*Definition 2. Replacement of the covariance matrix of measuremenst $K_n$ of the observed random field by matrix $K_n^m$ is called the discrete Markov approximation of connectivity m or just the discrete Markov approximation of a random field (n - number of measurements of the field, m - connectivity rate of Markov field; $n \geq m$ ).*

At the same time, observed (approximated) random field can be either a non-Markov, or Markov field of higher connectivity than approximating Markov field. There can be offered a lot of different ways of Discrete Markov Approximation (DMA) of random fields. Below there is shown one possible way of such approximation, which guarantees (for the tasks related to GLSE estimation) that the resulting estimates are:

- no worse than weighted GLS estimates, at connectivity of approximating Markov equal to zero (*m* = 0);
- coinciding with BLUE at connectivity of approximating field equal to *n*-1 (*m* = *n*-1), where *n* - the number of measurements.

Such approximation actually means that the measurements of non-Markov field are replaced by Markov field measurements of connectivity *m*. This makes it possible to use cost-effective computational schemes, used for Markov fields.

## IV.  ALGORITHM OF DISCRETE MARKOV APPROXIMATION (DMA)

DMA algorithm, proposed here, consists of two steps. Below it is shown, taking into account the symmetry of $K_n$ matrix and results given in Appendix A.

1) Half band with width *m* +*1* is "cut" from the original matrix $K_n$, that is the main diagonal and *m* parallel diagonals of its top or bottom are selected $0 \leq m \leq n-1$. Using elements of selected half band and formula (16), vectors $\Gamma_i$ $(i = \overline{1, n-1})$ are computed. Thus, $mn - m(m+1)/2$ coefficients $\gamma_{ij}$ are found, where the index *i* varies from 1 to n-1, and the index *j* ranges from 1 to m at $i \geq m$ and from 1 to i at $i < m$.

2) The elements of $K_n$ matrix outside of band with width $2m+1$ are replaced by new elements, calculated using formula

$$k_{ij} = \begin{cases} \Gamma_{i-1}^T \mathbf{k}_{[i-1],j} & j > i \\ \mathbf{k}_{i,[j-1]}^T \Gamma_{i,j-1} & j < i \end{cases} (i, j = \overline{1, n}). \quad (7)$$

Thus, a matrix $K_n^m$ is obtained, in which the elements within the band with width $2m+1$ coincide with the corresponding elements of the original matrix $K_n$, and the elements outside of the band are calculated with formula (7) with the help of elements situated inside the band.

*Note 1.* In many cases it is sufficient to perform the first step only, because for many purposes it is not necessary to know the off-diagonal elements $K_n^m$. For example, inversion of the matrix $K_n^m$ just needs knowledge of vectors, $\Gamma_i (i = \overline{1, n-1})$, found at the first step, and diagonal elements $k_{ii} (i = \overline{1, n})$. At the same time, the elements can be easily calculated with the help of formulas (18)-(20).

*Note 2.* If *m* = 0, then matrix $K_n^m = K_n^0$ is diagonal with diagonal elements coinciding with the diagonal elements of matrix $K_n$. Since the diagonal elements of $K_n$ are the dispersions of the field at the measuring points, estimates of GLSE coincide with $W_n = K_n^{-1} = K_n^{-0}$ weighted LSE estimates.

If *m* = 1, then in matrices $K_n$ and $K_n^m = K_n^1$ elements of three diagonals (main and two parallel neighbour diagonals) will coincide, if *m* = 2, then elements of five diagonals of matrices $K_n^m = K_n^2$ will be the same, etc., if *m* = *n-1*, then matrix $K_n^{n-1}$ completely coincides with the matrix $K_n$. Thus, changing *m* from 0 to *n-1* ,approximation accuracy of matrix $K_n$ can be varied by matrix $K_n^m$ .

***Definition 3.*** *The covariance matrix $K_n^{\,m}$, obtained by applying the DMA algorithm to the matrix $K_n$, will be named covariance matrix type $K_n^{\,m}$, adjoint with matrix $K_n$.*

*Note 3.* The approximation of the observed field by adjoint Markov field with connectivity $m = 0$ can be interpreted as discrete replacement of field at the measuring points by the *white noise* with dispersion, which coincides with the dispersion of the original field. Connectivity value $m = 1$ means approximation of the field at the measuring points by simple (singly connected ) Markov field, etc., the value of $m = n\ \text{-}1$ means approximation of the original field by $(n\text{-}1)$ - connected Markov field, coinciding at measuring points with the initial field . In the latter case, the covariance matrices $K_n$ and $K_n^{\,m}$, accordingly, of the initial and approximating Markov fields are the same.

To investigate issues of accuracy of GLSE estimates obtained using the DMA algorithm, there have been carried large series of numerical experiments (computer simulation) with the covariance functions of non-Markov processes, which are widely used in the practice of engineering studies. The results of these experiments are presented in the following section.

## V. Computer simulation of the accuracy of the DMA algorithm for the problem of parametric identification of mathematical expectation of the random process

To assess accuracy of the DMA algorithm, the problem, given above in expressions (1) - (6) for a random field with one-dimensional argument $t$ (random process), has been considered.

Let us assume that in points $T_n = \{t_1 < t_2 < \ldots < t_n \mid t_i \in T\}$ measurements are made, aiming at calculating parameter **B** estimates on the basis of these measurements. As parameter **B** estimates there were GLSE estimates $\mathbf{B}_n^{GLSE}$ (see (2) in section 2) with the weight matrix $W_n = A_n^{-1}$, where $A_n$ - positive definite matrix. In computer modeling the matrices of the following type were taken as $A_n$:

1)  $K_n$ - the covariance matrix of measurements of the observed process $Z(t)$ at points $T_n$;

2)  $K_n^{\,m}$ - matrix adjoint with $K_n$ (value of $m$ for the adjoint matrix varied from 1 to 5);

3)  $I_n$ - identity matrix of order $n$.

In the first case, the estimates of GLSE coincide with BLUE of the observed process, in the second case, these are estimates of GLSE for approximated process and, in the latter case, estimates of GLSE coincide with ordinary LSE estimates. The dispersion matrix of GLSE estimates of the parameters **B** considered for weight matrix has the following form (see formulas (3) and (6) of section 2):

$$1)\ D_n^{(1)} = \left(F_n K_n^{-1} F_n^T\right)^{-1}; \tag{8}$$

$$2)\ D_n^{(2)} = D_n^{(1)} F_n K_n^{-m} K_n K_n^{-m} F_n^T D_n^{(1)}; \tag{9}$$

$$3)\ D_n^{(3)} = \left(F_n F_n^T\right)^{-1} F_n K_n F_n^T \left(F_n F_n^T\right)^{-1}. \tag{10}$$

Comparing the matrices $D_n^{(1)}$, $D_n^{(2)}$ and $D_n^{(3)}$ (or, if $p > 1$, where $p$ - the number of unknown parameters, predetermined functionals $\Psi$ of these matrices) for various combinations of functions $k(s,t)$ and $f(t)$ and different schemes of measurements it is possible to obtain empirical evaluation of the accuracy of CM-approximation.

The determinant of the matrix (det) and the trace (tr) of dispersion matrix were chosen as the functional $\Psi$. The stationary functions of the following forms (see, e.g., [10]) were used as the covariance function of the field $Z(t)$:

$$1)\ k(\tau) = \exp(-\alpha_1 \mid \tau \mid)\cos\alpha_2\tau;$$

$$2)\ k(\tau) = \exp(-\alpha_1\tau^2);$$

$$3)\ k(\tau) = \exp(-\alpha_1\tau^2)\cos\alpha_2\tau.$$

Functions 1-3 for different values of the coefficients $\alpha_i$ are shown at Fig. 1- 3. It should be mentioned that function 1 represents covariance function of the first component of two-dimensional Markov field, functions 2 and 3 are the covariance functions of non- Markov processes.

Polynomials of 0, 1, 2 and 3 degrees were selected as functions $\mathbf{f}(t)$, which were the basis for calculating matrix $F_n$ in (8) - (10), i.e. mathematical expectation models had the following form:

$$1)\ \eta(t) = \beta_1;$$

$$2)\ \eta(t) = \beta_1 + \beta_2 t;$$

$$3)\ \eta(t) = \beta_1 + \beta_2 t + \beta_3 t^2;$$

$$4)\ \eta(t) = \beta_1 + \beta_2 t + \beta_3 t^2 + \beta_4 t^3.$$

There were also considered regression functions in the form of a Gaussian curve, when $\eta(t) = \beta_1 \exp\left(-\theta(t - t_0)^2\right)$.

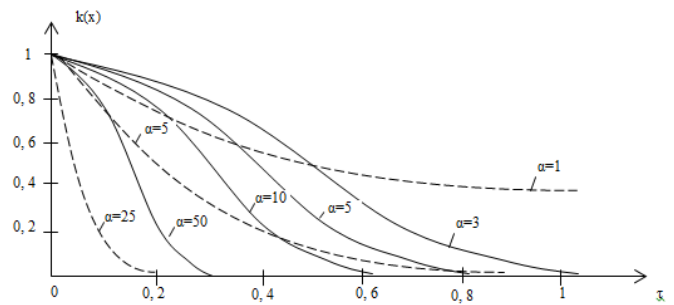

Fig. 1. Covariance functions $k(\tau) = \exp(-\alpha\tau^2)$ (solid lines) and $k(\tau) = \exp(-\alpha \mid \tau \mid)$ (dot-dashed lines) at different values of $\alpha$

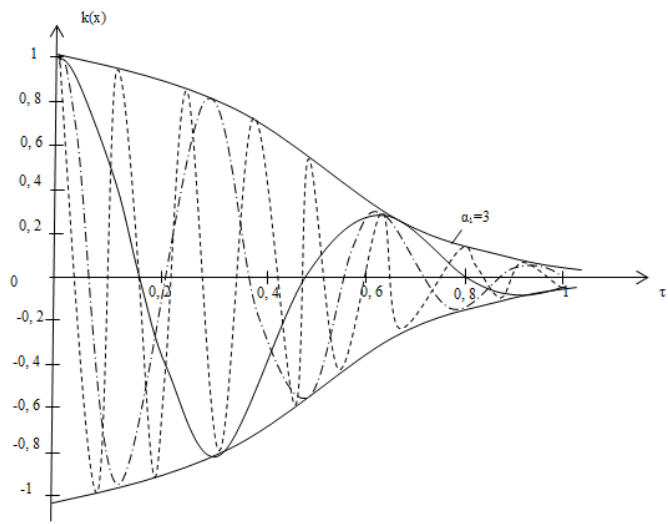

Fig. 2. Function $k(\tau) = \exp(-3\tau^2)\cos\alpha_2\tau$ at different values of $\alpha_2$
($\alpha_2 = 10$ - solid line, $\alpha_2 = 20$ - dot-dashed line and $\alpha_2 = 50$ - dotted line)

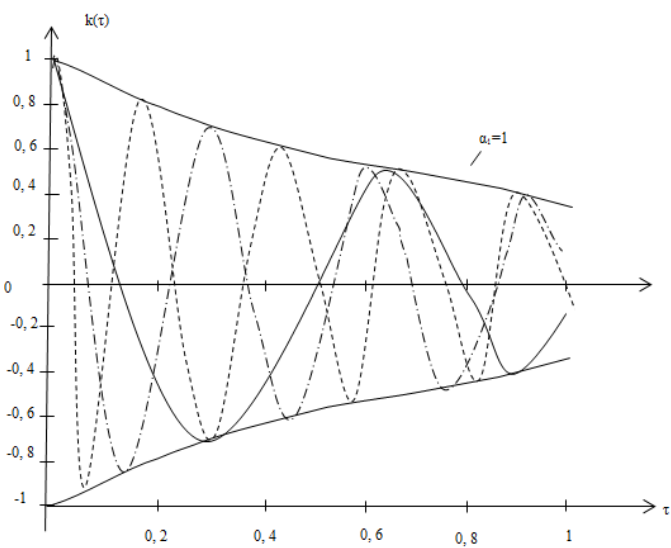

Fig. 3. The function $k(\tau) = \exp(-|\tau|)\cos\alpha_2\tau$ for different values $\alpha_2$
($\alpha_2 = 10$ - solid line, $\alpha_2 = 20$ – dot-dashed and $\alpha_2 = 40$ - dotted line)

In the experiments, the parameters $\alpha_1 \div \alpha_3$ and the parameter $\theta$ were varied over a wide range, permitting to cover a large range of real experimental situations.

As $T_n$ there were used 10- and 18-point measurements schemes with equidistant and unevenly spaced points. Part of computer simulation results are given in Appendix B.

Fig. 4 - 7 show the most typical functions $\det D_n^i \ (i = \overline{1,3})$, depending on the value of $m$ of approximating matrix $K_n^m$.

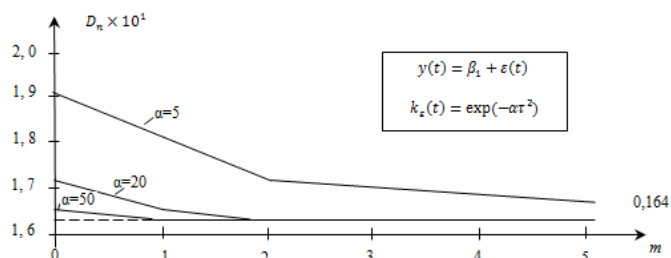

Fig. 4. Dependence of dispersion $D_n$ of parameter $\beta_1$ estimation on the width of half band $m$ of approximating matrix $K_n^m$.

*Notes. 1.* The value of $D_n$ at $m = n - 1$ corresponds to $D_n^{BLUE}$ - dispersion BLUE of parameter $\beta_1$, and corresponds to $D_n^{LSME}$ - dispersion of LSE estimate.

2. For clarity, the curve corresponding to $\alpha = 20$, is multiplied by a factor of 1.71, and the curve corresponding to $\alpha = 50$, by a factor of 2.43.

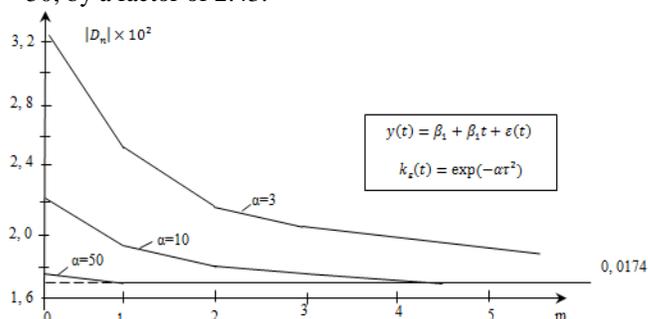

Fig. 5. Dependence of $\det D_n$ on the width of half band $m$ of approximating $K_n^m$.

*Notes. 1.* Value $\det D_n$ at $m = 0$ corresponds to $\det D_n^{LSME}$.

2. The horizontal line corresponds to $\det D_n^{BLUE}$.

3. The curve, corresponding to $\alpha = 10$, is multiplied by a factor of 1.881, and a curve, corresponding to $\alpha = 50$, by a factor of 5.073.

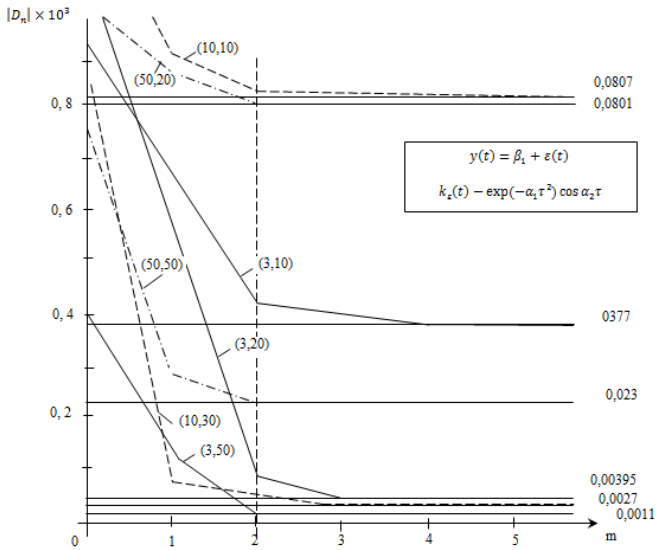

Fig. 6. Dependance of $\det D_n$ on the width of half band $m$ of approximating matrix $K_n^m$.

*Notes.* 1. See Notes 1 and 2 in Fig. 5.

2. The first number in parentheses corresponds to $\alpha_1$, the second – to $\alpha_2$

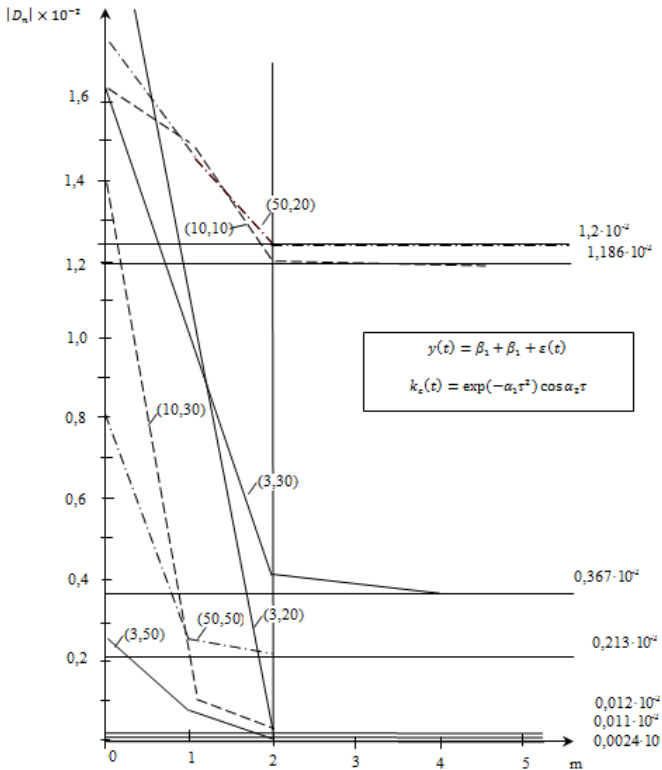

Fig. 7. Dependance of $\det D_n$ on the width of half band $m$ of approximating matrix $K_n^m$.

## VI. CONCLUSIONS

The results of computer simulation of DMA accuracy permit to make the following general conclusions about the possibility of using DMA in problems of GLSE estimation of random fields.

1) If measurements are weakly correlated in the area of the field observed, in other words, if the correlation interval is small relative to the interval of data removal, the use of optimal GLSE (BLUE) estimates provides light gain compared with conventional LSE estimates. In this case, it is advisable to abandon the optimal GLSE estimates in favor of the more computationally simple LSE estimates, even with well-known correlation properties of the process.

2) If observations are strongly correlated, particularly in cases where the covariance function contains a periodic component, the use of GLSE estimates provides a significant gain in accuracy of obtained estimates. For example, in some modeling situations, the value of the determinant of the dispersion matrix BLUE was dozens and hundreds of times less than the value of the determinant of the dispersion matrix of the LSE estimates (see. Appendix B). Obviously, in such situations the use of GLSE estimates with a weight matrix $W_n = K_n^{-1}$ is justified, despite the greater complexity of calculating them.

3) The results of numerical experiments show that in all model cases the approximation of the covariance measurements matrix by matrix $K_n^m$, even at small values of connectivity $m$ of approximating process allows to obtain GLSE estimates, whose dispersion matrix is virtually identical to the dispersion matrix BLUE. In experiments, such value of connectivity $m$ is not greater than 4.

4) Of particular interest is the fact, that for the covariance functions 1 and 3, containing a periodic component, the value of connectivity $m$, at which the dispersion matrix of GLSE estimates becomes scarcely distinguishable from the dispersion matrix BLUE, was equal 2, regardless of the model, of mathematical expectation of the process and of the experiment plan. Obviously, this is due to the fact that these functions are defined by two parameters and at connectivity of approximating Markov process equal to 2, all information about the covariance properties of the field may be obtained by the elements of the covariance measurements matrix, existing in half band with width $m$.

5) To sum up, it could be said that in case of strongly correlated measurements, use of optimal GLSE provides large gain in accuracy of obtained estimates. At the same time, to simplify the calculation of such estimates it could be recommended to use a discrete approximation process observed by Markov process of a small connectivity. As the results of numerical experiments show, at least for the polynomial regression functions and for covariance functions of the types 1-4, the discrete Markov approximation allows to achieve a good compromise between the accuracy of obtained estimates and the complexity of calculating them.

## VII. RESUME

1. A method for discrete approximation of the observed random field by $m$-connected Markov field is suggested, the method is called Discrete Markov Approximation (DMA) of a random field. The essence of the DMA algorithm is in specifically made replacement of covariance measurements matrix $K_n$ of the observed field by the covariance measurements matrix $K_n^m$ of $m$-connected Markov field.

2. The results of computer simulation of DMA accuracy for linear problems of parametric identification of the mathematical expectation of the field are presented.

3. The results of numerical experiments show that even at a low value connectivity $m$ of approximating Markov field, dispersion matrix of GLSE estimates with a weight matrix $W_n = K_n^{-m}$, it scarcely different from the dispersion matrix BLUE.

4. On this basis, it is recommended, in order to simplify the processing of strongly correlated measurements in problems of parametric identification, filtering and interpolation of random fields with GLSE to use DMA algorithm. This allows to achieve a good compromise between the accuracy of the estimates and the complexity of calculating them.

5. For weakly correlated measurements, as always, the best method of estimation, in terms of simplicity, is the usual LSE.

## APPENDIX A
## STRUCTURE OF COVARIANCE MEASUREMENTS MATRIX OF MARKOV PROCESS IN A WIDE SENSE (GAUSS-MARKOV RANDOM PROCESS)

### A. An ordinary-connected (simple) Markov process

Let the observed process Z (t) be a Markov process in a wide sense or Gauss-Markov random process (hereinafter everywhere it will be simply called "Markov process"). This means that the covariance function of Z (t) satisfies the condition of [8] (Dub, 1953):

$$k(s,t) = k(s,\tau)k(\tau,t)/k(\tau,\tau) \quad (s < \tau < t) \quad (11)$$

From Dub's theorem it also follows that the condition (11) is not only necessary but also sufficient, i.e. positive definite function $k(s,t)$ is the covariance function of the Markov process if and only if it satisfies condition (11) or, in other words, if the covariance function of a random process satisfies condition (11), then such process is a Markov process in a wide sense.

Let the values $\gamma i$ $(i = 1,2, ...)$ are defined as follows:

$$\gamma_i = k_{i,i+1}/k_{ii}, \quad (12)$$

where $k_{ij} = k(t_i, t_j)$ - values of covariance function of the process Z (t) at the points $t_i$ and $t_j$, $t_j > t_i$. Thus, $\gamma_i$ are covariance coefficients of the process at neighboring points coerced to the dispersion values at the points with a lower value of coordinates. For stationary random processes $\gamma_i = \rho_{i,i+1}$, i.e. $\gamma_i$ - correlation coefficients between neighboring measurement of the process.

Taking into account (11) and (12) the following theorem can be formulated.

*Theorem 1.* 1. Covariance matrix $K_n$ of measurement of the Markov process at the points $T_n = \{t_1 < t_2 < ... < t_n \mid t_i \in T\}$ could be represented as matrix (13), shown at the bottom of the page.

where $\Gamma_{ij} = \prod_{l=j}^{i} \gamma_l (i \geq j)$, $\Gamma_{ji} = \Gamma_{ij}(i \geq j)$, $(i, j = \overline{1, n-1})$ and $\gamma_i$ - is previously defined in (24).

2. The matrix is $K_n^{-1} = (K_n)^{-1}$ is tridiagonal.

3. The elements are defined by the following expressions:

$$\alpha_i = k_{ii} - \gamma_{i-1}^2 k_{i-1,i-1} \quad (i = \overline{2, n}), \quad \alpha_1 = k_{11},$$

$$\mu_i = k_{i+1,i+1} - \gamma_{i-1}^2 \gamma_i^2 k_{i-1,i-1} \quad (i = \overline{2, n-1}), \quad \mu_1 = k_{22}. \quad (14)$$

*Theorem 1' (reverse).* Any positive definite matrix of type $A_n^1$ is the covariance measurements matrix of a Markov process at points $T_n$.

Proving of theorems 1 and 1' are given in [16] (theorem 1 proving for more general case when the matrix $K_n = A_n^1$ is asymmetric, is also given in [17]).

*Corollary 1.* The matrix $K_n^1$ is completely determined by the elements of its two diagonals (the main and parallel to it diagonal above, or main and parallel one to it below). In other words, *the matrix $K_n^1$ depends only on n values of the dispersion of the process at the measuring points and (n - 1) covariance coefficients between the measurement points.*

### B. Multiply connected (m-connected) Markov process

Let Z(t) be an $m$-connected Markov process. This means that the covariance between the discrete measurements of the Z(t) process meets the condition

$$k(t_i, t_j) = \mathbf{k}_{i,[j-1]}^T K_m^{-1}[j-1]\mathbf{k}_{[j-1],j}, \quad (15)$$

$$K_n = K_n^1 = \begin{bmatrix} k_{11} & \Gamma_{11}k_{11} & \Gamma_{12}k_{12} & \cdot & \cdot & \Gamma_{1,n-1}k_{11} \\ \Gamma_{11}k_{11} & k_{22} & \Gamma_{22}k_{22} & \cdot & \cdot & \Gamma_{2,n-1}k_{22} \\ \Gamma_{21}k_{11} & \Gamma_{22}k_{22} & k_{33} & \cdot & & \cdot \\ \cdot & \cdot & \cdot & \cdot & & \cdot \\ \cdot & \cdot & & \cdot & k_{n-1,n-1} & \Gamma_{n-1,n-1}k_{n-1,n-1} \\ \Gamma_{n-1,1}k_{11} & \Gamma_{n-1,2}k_{22} & \cdot & \cdot & \Gamma_{n-1,n-1}k_{n-1,n-1} & k_{nn} \end{bmatrix}, \quad (13)$$

where $t_i < \cdots < t_{j-m} < t_{j-m+1} < \cdots < t_{j-1} < t_j$. Condition (15) can be obtained from (11) by transfer to matrix-vector notation. Thus, the symbol $[i]$ indicates that the index $i$ (or the variable $i$) changes from $i$-$m$ + 1 to $i$. If $i < m$, then index $i$ ranges from 1 to $i$

In (15) the following notations are used:

$\mathbf{k}_{i,[j-1]}^T = \left( k_{i,j-m}, k_{i,[j-m+1]}, \ldots, k_{i,j-1} \right)$ - $m$-dimensional

vector - $k(s,t)$ values row at the point $t_i$ and at the points

$T_m[j-1] = \left\{ t_{j-m}, t_{j-m+1}, \ldots, t_{j-1} \right\};$

$\mathbf{k}_{[j-1],j} = \left( k_{j-m,j}, k_{j-m+1,j}, \ldots, k_{j-1,j} \right)$ - $m$-dimensional

vector - $k(s,t)$ values column at the points $T_m[j-1]$ and at the point $t_i$ (in other words, $\mathbf{k}_{[i],l}$ is $m$-dimensional vector of measurements covariances at the point with the measurement at the point $t_l \in T_n$);

$K_m[i]$ - ($m$x$m$) covariance matrix of vector of values Z(t) at

the points $T_m[j-1]$, e.g. $K_m[j-1] =$

$[k(t_s, t_l)](s, l = \overline{j-m, j-1})$.

Graphic illustration of notations in formula (15) is shown at Figure 8.

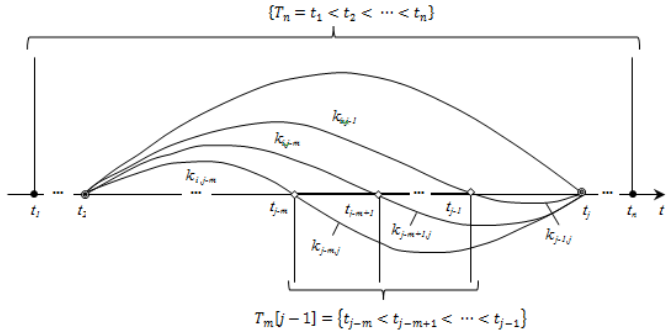

Fig. 8. Graphic illustration of formula (15)

Let the measurements of $m$-connected Markov process be made at the points $T_n$. Let $K_n$ be the covariance matrix of the measurements.

For matrix $K_n$ vector $\mathbf{k}_{[i],j} (i \geq j)$ in (A1.5) can be interpreted as a set of elements of $j$-th column from $i$-$m$+1 to $i$ at $i \geq m$ or from 1 to $i$ at $i < m$. Thus, the dimension of the vector $\mathbf{k}_{[i],j}$ will be equal $m$ at $i \geq m$ and will be equal $i$ at $i < m$.

Let us define (sub)matrices $K_m[i]$, which could be interpreted as diagonal submatrices of matrix $K_n$, located at the intersection of rows and columns of the same name with numbers from $i$-$m$+1 to $i$ at $i \geq m$ or from 1 to $i$ at $i < m$. Thus, the size of submatrix $K_m[i]$ will be equal $m$x$m$ at $i \geq m$ and $i$x$i$ at $i < m$.

Let the vectors $\Gamma_i (i = \overline{1, n-1})$ are defined as follows:

$$\Gamma_i = K_m^{-1}[i]\mathbf{k}_{[i],i+1} \,,\, (i = \overline{1, n-1}) \qquad (16)$$

Obviously, the dimension of the vector $\Gamma_i$ is equal to $m$ at $i \geq m$ and equal to $i$ at $i < m$.

Taking into account introduced notations the following theorem could be formulated.

*Theorem 2.* 1. The covariance matrix $K_n$ of measurements at points $T_n$ of m-connected Markov process could be represented in matrix form (17), shown at the bottom of the page.

2. Matrix $K_n^{-m} = \left[ K_n^m \right]^{-1}$, inverse to $K_n^m$ is a band type with half-width band equal to $m$.

3. The elements are defined as follows

$\alpha_i = k_{ii} - \mathbf{k}_{i,[i-1]}^T K_m^{-1}[i-1]\mathbf{k}_{[i-1],i} = k_{ii} - \mathbf{k}_{i,[i-1]}^T \Gamma_{i-1}$,

$(i = \overline{1, n})$, $\qquad (18)$

$c_{ii} = \frac{1}{\alpha_i} + \sum_{k=0}^{w} \frac{\gamma_{m-k,i+k}^2}{\alpha_{i+k+1}}$, $(i = \overline{1, n})$, $\qquad (19)$

$c_{i,i+k} = -\frac{\gamma_{i+k-1,m-k+1}}{\alpha_{i+k}} + \sum_{j=k}^{w} \frac{\gamma_{i+j,m-j}\gamma_{i+j,m+k-j}}{\alpha_{i+j+1}}$,

$c_{i+k,i} = c_{i,i+k}$, $(i = \overline{1, n-1}; k = \overline{1, m})$, $\qquad (20)$

where $w = m$-1 at $i \leq n - m$, and $w = n - i - 1$ at $i > n - m$.

$$K_n = K_n^m = \begin{array}{c} \\ 1 \\ 2 \\ 3 \\ \vdots \\ n \end{array} \begin{array}{cccccc} 1 & 2 & 3 & \cdots & n \\ \left\| \begin{array}{ccccc} k_{11} & \mathbf{k}_{1,[1]}^T\Gamma_1 & \mathbf{k}_{1,[2]}^T\Gamma_2 & \cdots & \mathbf{k}_{1,[n-1]}^T\Gamma_{n-1} \\ \Gamma_1^T\mathbf{k}_{[1],1} & k_{22} & \mathbf{k}_{2,[2]}^T\Gamma_2 & \cdots & \mathbf{k}_{2,[n-1]}^T\Gamma_{n-1} \\ \Gamma_2^T\mathbf{k}_{[2],1} & \Gamma_2^T\mathbf{k}_{[2],2} & k_{33} & & \vdots \\ \vdots & \vdots & & k_{n-1,n-1} & \mathbf{k}_{n-1,[n-1]}^T\Gamma_{n-1} \\ \Gamma_{n-1}^T\mathbf{k}_{[n-1],1} & \Gamma_{n-1}^T\mathbf{k}_{[n-1],2} & \cdots & \Gamma_{n-1}^T\mathbf{k}_{[n-1],n-1} & k_{nn} \end{array} \right\| \end{array} \qquad (17)$$

*Corollary 2.* The matrix $K_n^m$ is completely determined by its elements locating inside the band with the width of $2m+1$, where $m$ - the connectivity coefficient of multiconnected Markov process.

**Note**. If in (19) and (20) the calculated upper limit value becomes smaller than the lower one, then the summation is not executed, i.e. at $j,k > w$ and $j,k > m$ the second term in the right side of the formulas above is identically equal to 0.

## APPENDIX B
## RESULTS OF COMPUTER SIMULATION OF THE ACCURACY OF THE DISCRETE MARKOV APPROXIMATION FOR THE PROBLEM OF LINEAR PARAMETRIC IDENTIFICATION

Tables 1-12 shows the values of $\det D$ (numerator) and $trD$ (denominator) for different values $m$ of half band of approximating matrix $K_n^m$, where $D = D_n^{GLSE}$ - the dispersion matrix of GLSE estimates of the parameters $\beta_i$ (see, Section 2) obtained at the number of measurements $n = 16$.

$\Psi_1$ - dispersion of OLS estimate; $\Psi_2$ - dispersion BLUE.

The observed process is defined by a model $z(t) = \sum_{i=1}^p \beta_i f_i(t) + \xi(t)$, where $\beta_i$ - the estimated parameters; $f_i(t)$ - polynomials of order $p$ ($p = \overline{0,3}$); $\xi(t)$ - measurements noise with known covariance function $k_\xi(s,t) = k_\xi(\tau)$; $\tau = s - t$.

### TABLE 1
$Z(t) = \beta_1 + \xi(t)$ ; $k_\xi(\tau) = \exp(-\alpha\tau^2)$.

| $\alpha$ \ $m$ | 0 | 1 | 2 | 3 | 5 | n-1 | $\Psi_1/\Psi_2$ |
|---|---|---|---|---|---|---|---|
| 5 | 1.921 | - | 1.737 | 1.703 | 1.660 | 1.640 | 1.169 |
| 20 | 1.010 | 0.972 | 0.961 | 0.959 | 0.959 | 0.959 | 1.053 |
| 50 | 0.681 | 0.677 | 0.677 | 0.677 | 0.677 | 0.677 | 1.006 |

Notes. 1. Here, $p = 0$, and $\det D = trD_n = \sigma_n^2(\hat\beta_1)$.
2. The values of D are multiplied by $10^4$.

### TABLE 2
$Z(t) = \beta_1 + \beta_2 t + \xi(t)$ ; $k_\xi(\tau) = \exp(-\alpha\tau^2)$.

| $\alpha$ \ $m$ | 0 | 1 | 2 | 3 | 5 | n-1 | $\Psi_1/\Psi_2$ |
|---|---|---|---|---|---|---|---|
| 3 | 3.28 / 0.6031 | 2.58 / 0.4826 | 2.10 / 0.4720 | 2.10 / 0.4560 | 1.95 / 0.4510 | 1.74 / 0.4320 | 1.89 / 1.397 |
| 10 | 1.19 / 0.3699 | 1.00 / 0.3197 | 0.952 / 0.3140 | 0.935 / 0.3110 | 0.927 / 0.3100 | 0.925 / 0.3100 | 1.286 / 1.193 |
| 50 | 0.348 / 0.220 | 0.343 / 0.218 | 0.343 / 0.218 | 0.343 / 0.218 | 0.343 / 0.218 | 0.343 / 0.218 | 1.015 / 1.010 |

Note. The values of $\det D$ are multiplied by $10^2$.

### TABLE 3
$Z(t) = \beta_1 + \beta_2 t + \beta_3 t^2 + \xi(t)$ ; $k_\xi(\tau) = \exp(-\alpha\tau^2)$.

| $\alpha$ \ $m$ | 0 | 1 | 2 | 3 | 5 | n-1 | $\Psi_1/\Psi_2$ |
|---|---|---|---|---|---|---|---|
| 3 | 2.38 / 1.10 | 1.59 / 0.96 | 1.35 / 0.92 | 1.26 / 0.92 | 1.12 / 0.90 | 0.90 / 0.84 | 2.64 / 1.31 |
| 10 | 0.648 / 0.3699 | 0.543 / 0.694 | 0.512 / 0.686 | 0.506 / 0.684 | 0.500 / 0.682 | 0.500 / 0.682 | 1.30 / 1.095 |
| 50 | 0.228 / 0.622 | 0.189 / 0.579 | 0.183 / 0.563 | 0.180 / 0.568 | 0.182 / 0.568 | 0.182 / 0.568 | 1.25 / 1.095 |

Note. The values of $\det D$ are multiplied by $10^3$.

### TABLE 4
$Z(t) = \beta_1 + \beta_2 t + \beta_3 t^2 + \beta_4 t^3 + \xi(t)$ ; $k_\xi(\tau) = \exp(-\alpha\tau^2)$.

| $\alpha$ \ $m$ | 0 | 1 | 2 | 3 | 5 | n-1 | $\Psi_1/\Psi_2$ |
|---|---|---|---|---|---|---|---|
| 3 | 1.247 / 2.269 | 0.738 / 1.961 | 0.6405 / 1.939 | 0.500 / 1.869 | 0.393 / 1.763 | 0.2875 / 1.670 | 4.337 / 1.36 |
| 10 | 0.0409 / 5.43 | 0.0385 / 5.62 | 0.0239 / 4.82 | 0.0154 / 4.36 | 0.0149 / 4.33 | 0.0149 / 4.33 | 2.745 / 1.264 |

Note. The values of $\det D$ are multiplied by $10^4$.

### TABLE 5
$Z(t) = \beta_1 + \xi(t)$ ; $k_\xi(\tau) = \exp(-\alpha_1\tau^2)\cos\alpha_2\tau$ .

| $\alpha_1$ | $\alpha_2$ | 0 | 1 | 2 | 3 | 4 | n-1 | $\Psi_1/\Psi_2$ |
|---|---|---|---|---|---|---|---|---|
| 3 | 10 | 0.925 | 0.666 | 0.420 | 0.397 | 0.382 | 0.377 | 2.45 |
|  | 20 | 1.028 | 0.6337 | 0.0767 | 0.0443 | 0.0410 | 0.0395 | 26.03 |
|  | 50 | 0.4006 | 0.1299 | 0.0144 | 0.0126 | 0.1216 | 0.0113 | 35.45 |
| 10 | 10 | 1.18 | 0.9195 | 0.8151 | 0.8067 | 0.8066 | 0.8066 | 1.463 |
|  | 30 | 0.8401 | 0.0754 | 0.0382 | 0.0277 | 0.0274 | 0.0273 | 30.72 |
| 50 | 20 | 0.9929 | 0.8632 | 0.8021 | 0.8012 | 0.8012 | 0.8012 | 1.239 |
|  | 50 | 0.7577 | 0.2956 | 0.2320 | 0.2315 | 0.2315 | 0.2315 | 3.27 |

Notes. 1. Since p = 0, $\det D = trD_n = \sigma_n^2(\hat\beta_1)$
2. The values of $D$ are multiplied by $10^1$.

### TABLE 6
$Z(t) = \beta_1 + \beta_2 t + \xi(t)$ ; $k_\xi(\tau) = \exp(-\alpha_1\tau^2)\cos\alpha_2\tau$ .

| $\alpha_1$ | $\alpha_2$ | 0 | 1 | 2 | 3 | 5 | n-1 | $\Psi_1/\Psi_2$ |
|---|---|---|---|---|---|---|---|---|
| 3 | 10 | 1.62 / 0.3446 | 1.01 / 0.2981 | 0.405 / 0.1949 | 0.390 / 0.1930 | 0.370 / 0.1900 | 0.367 / 0.1896 | 4.41 / 1.82 |
|  | 20 | 2.17 / 0.3875 | 1.03 / 0.297 | 0.018 / 0.080 | 0.013 / 0.063 | 0.0107 / 0.066 | 0.0106 / 0.066 | 203.95 / 5.90 |
|  | 50 | 0.259 / 0.240 | 0.079 / 0.171 | 0.0033 / 0.069 | 0.0028 / 0.064 | 0.0024 / 0.060 | 0.00235 / 0.059 | 110.03 / 4.07 |
| 10 | 10 | 1.63 / 0.3451 | 1.505 / 0.3169 | 1.20 / 0.2914 | 1.19 / 0.2901 | 1.186 / 0.2900 | 1.186 / 0.2900 | 1.374 / 1.190 |
|  | 30 | 1.40 / 0.339 | 0.124 / 0.224 | 0.023 / 0.083 | 0.0122 / 0.061 | 0.0120 / 0.0608 | 0.0120 / 0.0608 | 116.67 / 5.58 |
| 50 | 20 | 1.76 / 0.336 | 1.49 / 0.320 | 1.25 / 0.303 | 1.25 / 0.3028 | 1.25 / 0.3028 | 1.25 / 0.3028 | 1.408 / 1.11 |
|  | 50 | 0.791 / 0.295 | 0.252 / 0.268 | 0.214 / 0.265 | 0.213 / 0.265 | 0.213 / 0.265 | 0.213 / 0.265 | 3.71 / 1.113 |

Note. The values of $\det D$ are multiplied by $10^2$.

TABLE 7

$$Z(t) = \beta_1 + \beta_2 t + \beta_3 t^2 + \xi(t); \quad k_\xi(\tau) = \exp(-\alpha_1 \tau^2)\cos\alpha_2 \tau.$$

| $\alpha_1$ | $\alpha_2$ \ $m$ | 0 | 1 | 2 | 3 | 5 | n-1 | $\Psi_1/\Psi_2$ |
|---|---|---|---|---|---|---|---|---|
| 3 | 10 | 3.54 / 0.975 | 2.25 / 0.906 | 0.747 / 0.78 | 0.727 / 0.78 | 0.680 / 0.77 | 0.675 / 0.77 | 5.24 / 1.27 |
| | 20 | 3.22 / 0.818 | 1.2 / 0.772 | 0.019 / 0.622 | 0.015 / 0.614 | 0.012 / 0.607 | 0.0112 / 0.606 | 287.5 / 1.34 |
| | 50 | 0.408 / 0.798 | 0.106 / 0.683 | 0.0028 / 0.583 | 0.0035 / 0.583 | 0.0026 / 0.578 | 0.00198 / 0.578 | 206.06 / 1.38 |
| 10 | 10 | 5.82 / 1.034 | 4.05 / 0.973 | 3.00 / 0.910 | 2.94 / 0.908 | 2.94 / 0.908 | 2.94 / 0.908 | 1.98 / 1.14 |
| | 30 | 3.398 / 0.890 | 1.66 / 0.611 | 0.0271 / 0.560 | 0.0143 / 0.552 | 0.0142 / 0.552 | 0.0142 / 0.552 | 239.3 / 1.61 |
| 50 | 20 | 4.23 / 0.898 | 3.63 / 0.88 | 2.98 / 0.87 | 2.978 / 0.87 | 2.978 / 0.87 | 2.978 / 0.87 | 1.42 / 1.032 |
| | 50 | 2.20 / 1.015 | 0.714 / 0.996 | 0.60 / 0.990 | 0.60 / 0.990 | 0.60 / 0.990 | 0.60 / 0.990 | 3.67 / 1.025 |

Note. The values of $\det D$ are multiplied by $10^3$.

TABLE 8

| $b$ | $\alpha$ \ $m$ | 0 | 1 | 2 | 3 | 4 | n-1 | $\Psi_1/\Psi_2$ |
|---|---|---|---|---|---|---|---|---|
| 3 | 3 | 1.050 | 1.322 | 1.105 | 1.006 | 1.005 | - | 1.045 |
| | 10 | 0.744 | 0.836 | 0.779 | 0.730 | 0.726 | 0.715 | 1.04 |
| | 50 | 0.4505 | 0.397 | 0.375 | 0.373 | 0.373 | 0.373 | 1.21 |
| 20 | 3 | 1.345 | 0.300 | 0.072 | 0.016 | 0.0068 | - | ~200 |

TABLE 9

$$Z(t) = \beta_1 \exp(-10t^2) + \xi(t); \quad k_\xi(\tau) = \exp(-\alpha_1 \tau^2)\cos\alpha_2 \tau.$$

| $\alpha_1$ | $\alpha_2$ \ $m$ | 0 | 1 | 2 | 3 | 4 | n-1 | $\Psi_1/\Psi_2$ |
|---|---|---|---|---|---|---|---|---|
| 3 | 10 | 0.360 | 0.111 | 0.039 | 0.027 | 0.025 | 0.022 | 16.36 |
| | 50 | 0.0323 | 0.015 | 0.0012 | 0.00047 | 0.00017 | 0.00013 | 248.46 |
| 10 | 10 | 0.4835 | 0.307 | 0.220 | 0.209 | 0.204 | 0.203 | 2.38 |
| | 30 | 0.2185 | 0.156 | 0.047 | 0.028 | 0.027 | 0.025 | 8.7 |

TABLE 10

$$Z(t) = \beta_1 \exp(-t^2) + \xi(t); \quad k_\xi(\tau) = \exp(-10\tau^2)\cos\alpha_2 \tau.$$

| $\alpha_2$ \ $m$ | 0 | 1 | 2 | 3 | 4 | n-1 | $\Psi_1/\Psi_2$ |
|---|---|---|---|---|---|---|---|
| 20 | 0.1714 | 0.0281 | 0.0163 | 0.0103 | 0.0094 | 0.0093 | 18.43 |
| 50 | 0.082 | 0.013 | 0.0046 | 0.0037 | 0.0033 | 0.00325 | 25.23 |

TABLE 11

$$Z(t) = \beta_1 \exp(-20t^2) + \xi(t); \quad k_\xi(\tau) = \exp(-\alpha_1 |\tau|)\cos\alpha_2 \tau$$

| $\alpha_1$ | $\alpha_2$ \ $m$ | 0 | 1 | 2 | 3 | 4 | n-1 | $\Psi_1/\Psi_2$ |
|---|---|---|---|---|---|---|---|---|
| 1 | 10 | 0.617 | 0.284 | 0.179 | 0.169 | 0.168 | 0.1675 | 3.68 |
| | 30 | 0.263 | 0.115 | 0.0551 | 0.053 | 0.053 | 0.053 | 4.97 |
| 5 | 10 | 0.659 | 0.543 | 0.525 | 0.523 | 0.523 | 0.523 | 1.26 |
| | 40 | 0.1204 | 0.117 | 0.1094 | 0.1093 | 0.1093 | 0.1093 | 1.102 |
| 25 | 20 | 0.4201 | 0.417 | 0.416 | 0.416 | 0.416 | 0.416 | 1.001 |
| | 50 | 0.2611 | 0.256 | 0.256 | 0.256 | 0.256 | 0.2559 | 1.02 |

TABLE 12

$$Z(t) = \beta_1 + \xi(t); \quad k_\xi(\tau) = \exp(-\alpha_1 |\tau|)\cos\alpha_2 \tau.$$

| $\alpha_1$ | $\alpha_2$ \ $M$ | 0 | 1 | 2 | 3 | 4 | n-1 | $\Psi_1/\Psi_2$ |
|---|---|---|---|---|---|---|---|---|
| 1 | 10 | 0.06095 | 0.0593 | 0.0358 | 0.0346 | 0.0345 | 0.03448 | 1.77 |
| | 30 | 0.0736 | 0.0479 | 0.0246 | 0.0239 | 0.02385 | 0.02385 | 3.085 |
| 5 | 10 | 0.1054 | 0.0949 | 0.0928 | 0.0927 | 0.0927 | 0.0927 | 1.24 |
| | 40 | 0.0997 | 0.0709 | 0.0622 | 0.0622 | 0.0622 | 0.06216 | 1.604 |
| 25 | 20 | 0.1044 | 0.1035 | 0.1035 | 0.1035 | 0.1035 | 0.1035 | 1.009 |

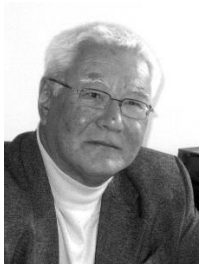

**Ulan N. Brimkulov** was born in Toktogul City, Kyrgyzstan. He received diploma of engineering in Automatic and Remote control in 1972, Ph.D (Candidate of Tech. Science) degree in 1978 in Tech. Cybernetics and Dr. of Science (Full Doctor) degree in scientific computing in 1992 from Moscow Power Institute (now Moscow Power University), Moscow, Russia.

He is presently a Professor of Computer Engineering Department and dean of Engineering Faculty of Kygyz-Turkish Manas University (Bishkek, Kyrgyzstan). From 1972 to 1992 he was an Assistant Professor, Associate Professor, Professor and Head of Computer Science Department in Frunze Polytechnic Institute (now Kyrgyz State Technical University, Bishkek, Kyrgyzstan). His research interests include design and analysis of experiments in random fields investigations. He has over 150 published papers (mainly in Russian), books and textbooks, including the book: U.N. Brimkulov, G.K. Krug, and V.L. Savanov. Design of experiments in investigating random fields and processes. Moscow: Nauka, 1986 (in Russian), and paper: U.N. Brimkulov, G.K. Krug, and V.L. Savanov, "Numerical construction of exact experimental designs when the measurements are correlated," Zavodskaya Laboratoria (Industrial Laboratory), Moscow. vol. 46, no. 5, pp. 475-480, 1980.

He is Corresponding Member of Kyrgyz Republic National Academy of Science, Member of: International Engineering Academy (Moscow); International Informatization Academy; Academy of Pedagogical and Social Sciences (Russia).

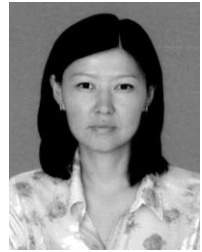

**Chinara Jumabaeva** was born in Kara-Balta, Kyrgyzstan. She received master degree of Computer Information Systems in 1999, Ph.D. (Candidate of Tech. Science) degree in 2013 in Automation and Control of Technological Processes and Manufactures, Bishkek, Kyrgyzstan.

She is a lecturer at the Kyrgyz-Turkish "Manas" University (Bishkek, Kyrgyzstan) since 2009.

Her research interest includes Computing Algorithms, Educational Management Information Systems, E-learning.

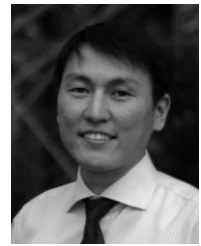

**Kasym Baryktabasov** received the B.S. degree in computer science from Kyrgyz National University, Bishkek, Kyrgyz Republic in 2002 and the M.S. degree in Computer Information Systems and Internet from the same university in 2004. He is currently pursuing the Ph.D. degree in computer engineering at Kyrgyz-Turkish "Manas" University, Bishkek, Kyrgyz Republic.

He is a lecturer with Kyrgyz-Turkish "Manas" University, Bishkek, Kyrgyz Republic since 2005. His research interest includes computing algorithms, knowledge management and e-government.

In 2010 he was engaged as a fellow at United Nations University International Institute for Software Technology Center for Electronic Governance to make research on Knowledge Management Infrastructure for Electronic Government and contribute to the ongoing projects.